\documentclass[aps,prl,reprint,superscriptaddress,longbibliography]{revtex4-1}
\usepackage{amsmath}
\usepackage{amsfonts}
\usepackage{color}
\usepackage{array}
\newcommand{\llangle}{\langle\!\langle}
\newcommand{\rrangle}{\rangle\!\rangle}
\newcommand{ \ba}{\begin{eqnarray}}
\newcommand{ \ea}{\end{eqnarray}}

\usepackage{epsf,latexsym,graphicx}
\usepackage{xr}
\usepackage{xcite}
\usepackage{pifont}
\externalcitedocument{supp}

\begin{document}

\title{ Collective mode across the BCS-BEC crossover in Holstein model }

\author{Tae-Ho Park}
\email{  thpark@skku.edu}
\affiliation{Department of Physics and Institute for Basic Science Research, Sungkyunkwan University, Suwon 16419,
Korea}

\author{Han-Yong Choi}
\email{ hychoi@skku.edu}
\affiliation{Department of Physics and Institute for Basic Science Research, Sungkyunkwan University, Suwon 16419,
Korea}
\affiliation{Asia Pacific Center for Theoretical Physics, Pohang 37673, Korea}

\date{\today}

\begin{abstract}

We investigate the emergence of the collective mode in the phonon spectra of the superconducting state within the Holstein model by varying the electron-phonon coupling. Using dynamical mean field theory (DMFT) combined with the numerical renormalization group (NRG) technique, we calculate the phonon spectra. In the superconducting state with a pairing gap ($\Delta_P$), the peak position of the collective mode ($\omega_{col}$) evolves from the Bardeen-Cooper-Schrieffer (BCS) regime, manifesting near $2\Delta_P$ and increasing with coupling, to the Bose-Einstein condensation (BEC) regime, where $\omega_{col}$ decreases with increasing coupling. The decrease of $\omega_{col}$ matches well with the reduction of superfluid stiffness, which originates from the increasing phase fluctuations of local pairs with coupling strength. In the crossover regime with intermediate coupling, $\omega_{col}$ aligns with the soft phonon mode ($\omega_s$) of the normal state and decreases with increasing coupling when $\omega_s < 2\Delta_P$. Additionally, comparing the collective mode weight to $\Delta_P$  suggests that the collective mode predominantly stems from U(1) gauge symmetry breaking across all coupling strengths.
\end{abstract}


\maketitle

$Introduction$\label{sec:intro}
--
The exploration of collective modes arising from fluctuations in both amplitude and phase of the order parameter due to spontaneous symmetry breaking has garnered significant attention in the fields of condensed matter and cold-atom systems.\cite{Poniatowski_CP2022,Pekker_AR2015,Shimano_AR2020,Leonard_Science2017,Endres_Nature2012} 
In the superconducting state, the formation of Cooper pairs results in the breaking of U(1) gauge symmetry, leading to elegant quantum mechanical behaviors, such as the uncertainty relation between particle number $n$ and phase factor $\theta$ as $\Delta n \cdot \Delta \theta \sim 1$ where $\Delta n$ and  $\Delta \theta$ are fluctuations of $n$ and $\theta$ respectively, and we take the Plank constant $\hbar$ as 1.\cite{Elion_Nature1994,Emery_Nature1995} Within the framework of the Bardeen-Cooper-Schrieffer (BCS) theory, the collective mode stemming from phase fluctuations, known as the Goldstone mode, couples to the longitudinal (diamagnetic) component of the electromagnetic response, giving rise to phenomena like the Meissner effect.\cite{Anderson_PR1958Meissner,Anderson_PR1958RPA,Nambu_PR1960} In the presence of long-range Coulomb interaction, this phase mode shifts up to the plasmon frequency, known as Anderson-Higgs mechanism.\cite{Anderson_PR1963,Higgs_PRL1964} On the other hand, in the strong coupling regime characterized by the Bose-Einstein-condensation (BEC) of bound pairs, the phase fluctuation is intensified, leading to the suppression of both phase coherence and superfluid stiffness.\cite{griffin_1994,randeria_2014,nozieres_1985,Leggett_1980} The significant effect of phase fluctuations poses challenges for studying phase modes, particularly in strong coupling scenarios, where traditional perturbative treatments like the Ginzburg-Landau free energy description may be inadequate.

While the evolution of collective modes across the BCS-BEC crossover has been extensively studied using the attractive Hubbard model within the random phase approximation (RPA) scheme,\cite{micnas_1990,Kostyrko_PRB1992,Belkhir_PRB1992,Belkhir_PRB1994,Sofo_PRB1992} the significant effect of phase fluctuations in the strong coupling regime remains unresolved. Recent theoretical analyses have explored the evolution of the amplitude mode, revealing its overdamped behavior in the BEC regime due to the tune away from Lorentz invariance and its concealment below a branch cut.\cite{Liu_PRA2016,Phan_PRB2023} However, in the BCS regime, the amplitude mode, though challenging to observe due to its lack of coupling to charge in linear response,\cite{Littlewood_PRL1981,Littlewood_PRB1982} has been detected in systems where superconductivity and charge density wave orders coexist,\cite{Sooryakumar_PRL1980,Measson_PRB2014,Grasset_PRL2019} and more recently, through nonlinear optical response techniques.\cite{Matsunaga_PRL2012,Matsunaga_PRL2013,Matsunaga_Science2014}
In contrast, the observation of the phase mode in condensed matter systems has been less prevalent, largely due to overlooked mechanisms. This letter proposes a mechanism for observing the phase mode in the strong coupling regime, highlighting its coupling to the soft phonon mode and its close association with superfluid stiffness, which suppresses phase coherence via local bound pairs. Consequently, the phase mode may fall within the observable energy range, well below the plasmon frequency. 
Notably, in a cold-atom system comprising a harmonically trapped gas of $^6 \textrm{Li}$ fermionic atoms, the evolution of the Bogoliubov-Anderson mode along the BCS-BEC crossover has been reported through measurements of density-density correlation via two-photon Bragg spectroscopy.\cite{Hoinka_NaturePhys2017} 

The BCS-BEC crossover in the attractive Hubbard model is characterized by a gradual transition from a potential-energy-driven (BCS) to a kinetic-energy-driven (BEC) pairing mechanism.\cite{toschi_2005,toschi_2005b,grag_2005,koga_2011,bauer_2009,bauer_2009b,Kyung_PRB2006} Although the charge susceptibility spectra of the attractive Hubbard model can exhibit variations in charge excitation along this crossover,\cite{bauer_2009,bauer_2009b,Kyung_PRB2006} the precise mechanism for the evolution of collective modes remains elusive. This is primarily because charge excitations are typically represented as broad peaks due to the instantaneous on-site Coulomb interaction, making it challenging to explicitly estimate phase fluctuations. However, the Holstein model, which incorporates explicit gauge bosons, offers a more comprehensive framework. It can capture dynamical effects such as gauge boson softening and retardation behavior, providing a means to estimate quantum fluctuations across weak to strong coupling regimes.\cite{Park2019prb}

{\it Model and methods }\label{sec:model}
--
The Holstein model is given by
\ba\label{eq:Hol}
{\cal H} &=& -\frac{t}{\sqrt{q}} \sum_{\langle i,j\rangle\sigma} c_{i\sigma}^\dag c_{j\sigma} 
 -\mu \sum_i n_i 
\nonumber \\
&+&\omega_0 \sum_i a_i^\dag a_i
+ g \sum_i  x_i \left( n_i -1 \right),
\ea
where $c_{i\sigma}$ and $a_i $ are the fermion of spin $\sigma$ and boson operators at the site $i$, and $\langle i,j\rangle$ implies the nearest neighbors with the coordination number $q$, and
$n_i =\sum_\sigma c_{i\sigma}^\dag c_{i\sigma}$ is the electron density operator at site $i$, and $\mu$ is the chemical potential. The electrons are coupled with the Einstein
phonon of frequency $\omega_0$ with the onsite coupling constant $g$ and $x_i=a_i^\dag +a_i$ is the displacement operator of the site $i$.
A useful way to grasp this model, in comparison with the attractive Hubbard model, is to integrate out the phonons and express the effective interaction between electrons as
 $ U_{eff} (\omega) = 2g^2 \omega_0/(\omega^2 -\omega_0^2 )$.\cite{Hirsch_PRB1983}
The derivation is provided in the Supplementary Material (SM I).\cite{supplement}
In the limit $\omega_0 \rightarrow \infty $, $ U_{eff} (\omega)\rightarrow -E_b$, where $E_b =2g^2/\omega_0$ is the bipolaron energy and can replace the electron-phonon coupling. $D=2t$ represents the half-bandwidth, with $D$  taken as the unit of energy in this context. 
We address the Holstein model in infinite dimensions by employing the dynamical mean-field theory (DMFT) in conjunction with the numerical renormalization group (NRG) technique.\cite{georges_1996,bulla_2008} This non-perturbative calculation enables the capture of low-frequency features in the phonon and density-density correlation spectra.
Here, we solve the Holstein model for the antiadiabatic case with $\omega_0 = 2.0$ $( > t)$ at half-filling ($\langle n\rangle=1$) at zero temperature.
The Cooper pairing (P) term, breaking the particle number symmetry, is implemented in the bath of the effective local Hamiltonian
\begin{subequations}
\begin{equation}\label{eq:loc}
\begin{split}
{\cal H}_{loc} = & \sum_{k,\sigma} \varepsilon_k c_{k\sigma}^\dag c_{k\sigma}
+ \sum_{k,\sigma} V_k \left( f_{\sigma}^\dag c_{k\sigma} +
c_{k\sigma}^\dag f_{\sigma} \right) \\
& -\sum_{k,\sigma} \Delta_k^{P} \left( c_{k\uparrow}^\dag c_{-k\downarrow}^\dag
+c_{-k\downarrow} c_{k\uparrow} \right)  + {\cal H}_{imp},
\end{split}
\end{equation}
\begin{equation}\label{eq:imp}
{\cal H}_{imp} =  \omega_0  a^\dag a - \mu n_f + g x( n_f -1 ),
\end{equation}
\end{subequations}
where $f_\sigma$ is the fermionic operator on the impurity site, $n_f=\sum_\sigma f_\sigma^\dag f_\sigma$. 
In the impurity solver for the DMFT self-consistent calculation, $\varepsilon_k$ represents the energy of the conduction electrons with momentum $k$, 
$V_k$ is the hybridization parameter,\cite{krishnamurthy_1980} and $\Delta_k^{P}$ is the Cooper pairing amplitude as effective medium parameters. 
We follow the Lanczos transformation for the NRG  calculation from Ref. \onlinecite{bauer_2009}.
We take the NRG parameters $\Lambda=1.6$, $N_S=1000$, $N_{ph}=10$ for numerical calculations.\cite{Park2019prb}

\begin{figure}
\begin{center}
\includegraphics[width=\linewidth]{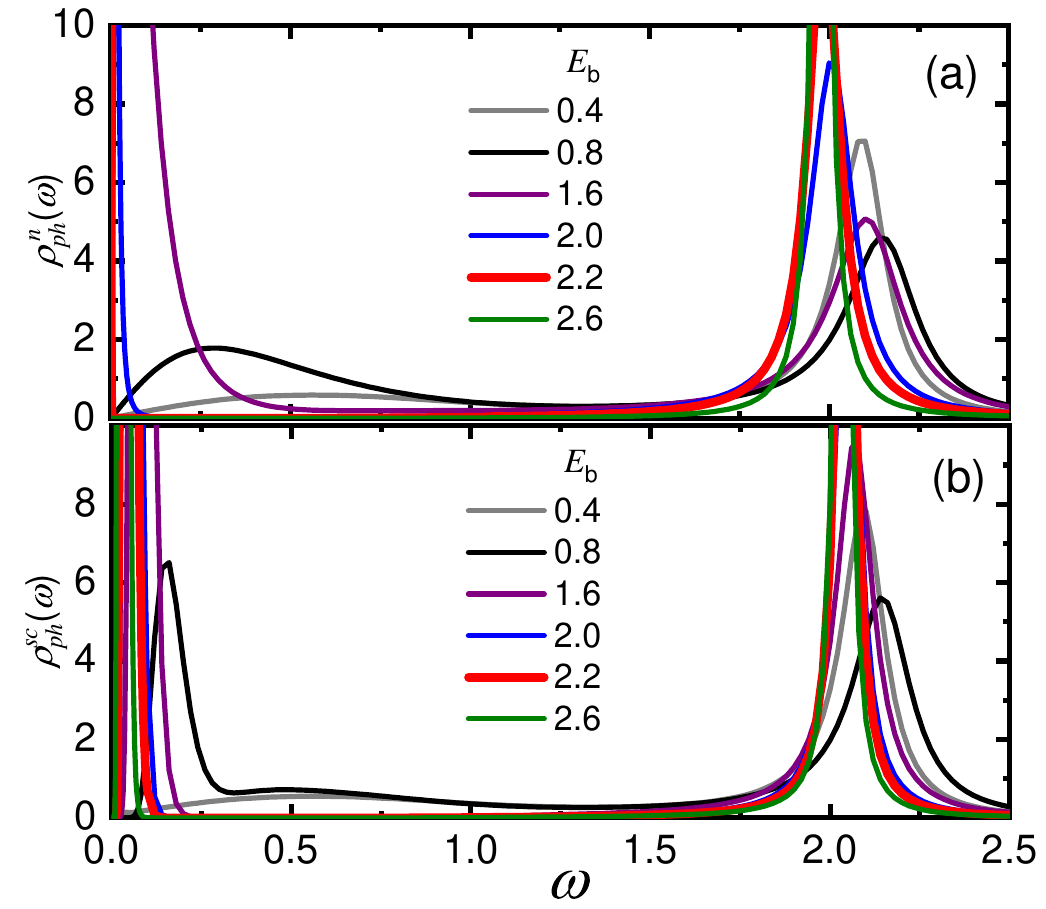}
 \caption{ Comparison of phonon spectra of the normal state $\rho_{ph}^n (\omega)$ (a) with the superconducting state $\rho_{ph}^{sc} (\omega)$ (b) for various $E_b$ from weak electron-phonon coupling (BCS regime) to strong coupling (BEC regime) through the crossover regime. The same color code is used for (a) and (b). Details are explained in the text.}
\label{fig:phDOS}
\end{center}
\end{figure}

{\it Phonon and density-density correlation spectra }\label{sec:resultsA}
--
We first compare the phonon spectra of the superconducting state with those of the normal state to elucidate the collective mode associated with the U(1) gauge symmetry breaking from the bosonic spectra.
The phonon Green's function, incorporating the self-energy correction is given by
\begin{equation}\label{eq:phGF}
 D^{-1}(\omega)= D_0^{-1}(\omega)-\Pi(\omega),
\end{equation}
where $D_0(\omega)=2\omega_0/(\omega^2-\omega_0^2)$ is the bare phonon Green's function and
$\Pi(\omega)=g\llangle n_f, x\rrangle_{\omega}/\llangle  x, x \rrangle_{\omega}$ is the phonon self-energy.
The double bracket $\llangle  \rrangle_{\omega}$ denotes the correlator as a function of frequency $\omega$. 
Details of the phonon Green's function and the correlator calculation are in the Supplemental Material (SM II).\cite{supplement}
Consequently, the phonon spectral function is obtained from $\rho_{ph}(\omega)= -1/\pi \textrm{Im} D(\omega)$.
Generally, in both normal and superconducting states, as the electron-phonon interaction is activated, the phonon spectra bifurcate into the low-frequency modes around zero and the high-frequency modes around the bare phonon frequency $\omega_0=2.0$, as shown in Fig. 1 (a) and (b) for normal ($\rho^{n}_{ph}(\omega)$) and superconducting ($\rho^{sc}_{ph}(\omega)$) states, respectively. 

The low-frequency modes exhibit softening towards lower frequencies due to the renormalization of the phonon mode with increasing the electron-phonon coupling $g$ (or $E_b$). 
The peak positions of collective mode $\omega_{col}$ and the soft phonon mode $\omega_s$ are obtained from
$d \rho_{ph}^{sc} (\omega) / d \omega \vert_{\omega=\omega_{col}} =0$  and $d \rho_{ph}^{n} (\omega) / d \omega \vert_{\omega=\omega_s} =0$, respectively, at the lowest frequency. Elaborate features of $\omega_{col}$ and $\omega_s$ will be discussed later.
On the other hand, with the increase in $E_b$, the high-frequency modes marginally harden towards frequencies higher than $\omega_0$, only to revert to the bare frequency with the significant development of the soft mode.

In the normal state, the metal-to-insulating transition occurs as a first-order transition, resulting in the coexistence of metallic and insulating states within the region between critical values $E_{c1}$ and $E_{c2}$, where the region for $E_b < E_{c1}$ corresponds to the metallic state and $E_b > E_{c2}$ for the insulating states.
This phase separation is usually determined by the electronic density of state in the DMFT calculation and is contingent upon the initial configurations.\cite{meyer_2002,capone_2003,jeon_2004,koller_2004}
Here, we obtained $E_{c1}=1.3$ from the insulating initial configuration and $E_{c2}=2.2$ from the metallic one. These critical values can also be discerned from the softening of the phonon mode associated with lattice instability.
Fig. 1(a) illustrates the normal state phonon spectra $\rho_{ph}^n(\omega)$ calculated from the metallic initial condition. Notably, at $E_b=2.2$ (red curve), pronounced phonon softening is observed, characterized by a sharp zero-frequency soft mode, while for the insulating state at $E_b=2.6$ (green curve), the soft mode vanishes, and the phonon mode reverts to the bare phonon frequency, where all the weight is restored on the single mode.

\begin{figure}
\begin{center}
\includegraphics[width=\linewidth]{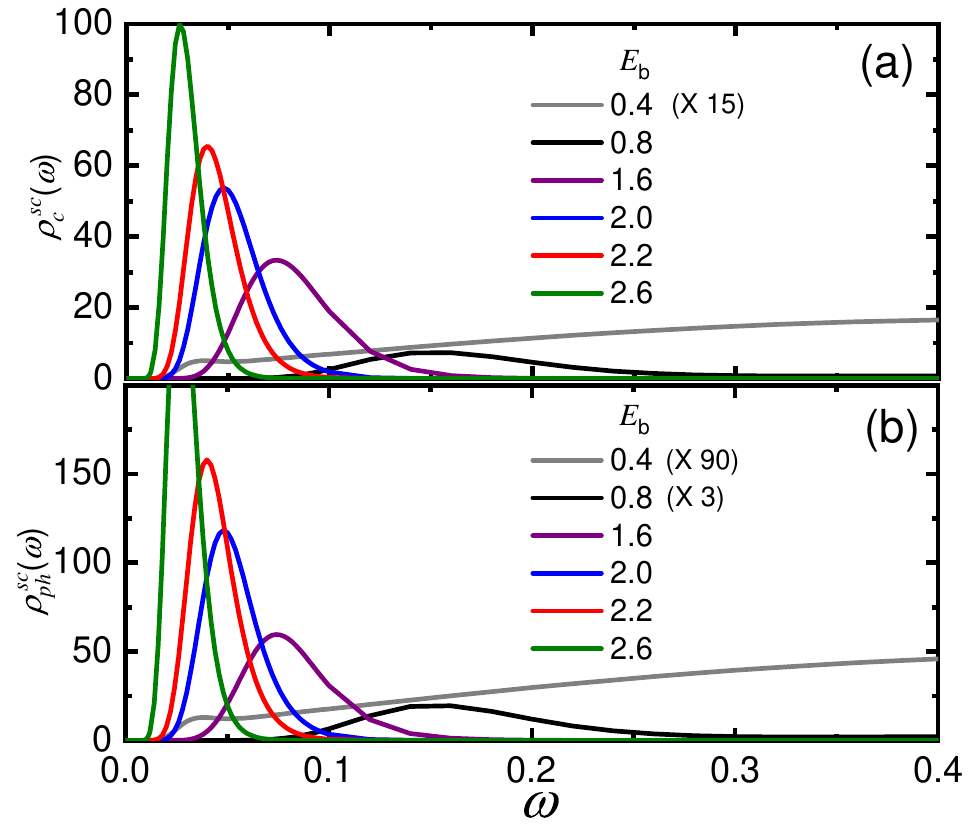}
 \caption{ Comparison of the density-density correlation spectra $\rho_{c}^{sc} (\omega)$ (a) to the phonon spectra $\rho_{ph}^{sc} (\omega)$ (b) in superconducting state for the same $E_b$ as in Fig.\ 1. Fig.\ 1(b) zooms in on panel (b) for the smaller frequency region. Small peaks at $E_b=0.4$ in (a) and $E_b=0.4$, $E_b=0.8$ in (b) are magnified by the numbers in the parenthesis to represent the peaks. }
\label{fig:ph_charge}
\end{center}
\end{figure}

\begin{figure}
\begin{center}
\includegraphics[scale=0.48]{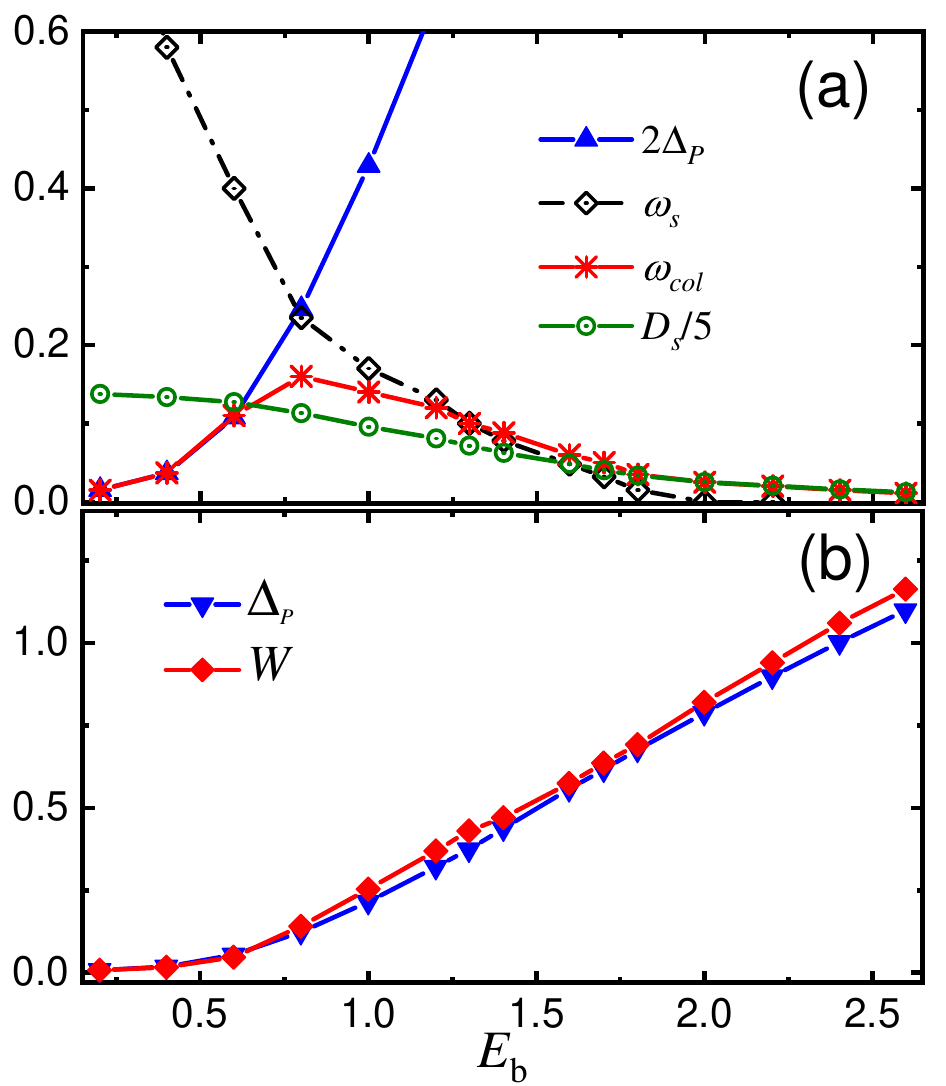}
 \caption{(a) Evolution of the collective mode $\omega_{col}$ (red (\ding{83}) points) as a function of  $E_b$ from weak electron-phonon coupling (BCS regime) to strong coupling (BEC regime) through the crossover regime for the Holstein model with the bare phonon frequency $\omega_0 = 2$ in the unit of $D$. Double of the pairing gap $2 \Delta_P$ (blue (\ding{115}) points), the soft phonon mode $\omega_s$ (black rhombuses) of the normal state and the superfluid stiffness $D_s$ divided by 5 (green circles) are presented together with the evolution of the collective mode. (b) Comparison of the pairing gap $\Delta_P$ (blue (\ding{116}) and the weight of the collective mode $W$ (red (\ding{117}) points). Details are explained and discussed in the text.}
\label{fig:evolution}
\end{center}
\end{figure}

However, in the superconducting state, characterized by a nonzero pairing parameter ($\Delta_P \neq 0$), the metal-insulator transition observed in the normal state transforms into a gradual crossover from the BCS regime for weak coupling to the BEC regime for strong coupling. Notably, in the superconducting state, a unique solution is obtained from the DMFT calculation regardless of the initial condition. In the BCS regime, the superconducting critical temperature $T_c$ is proportional to the pairing amplitude, which is determined by the potential energy. Meanwhile, in the BEC regime, $T_c$ is proportional to the superfluid stiffness ($D_s$) and $t^2/E_b$. Deviations between $T_c$ and $\Delta_P$ along the coupling strength are observed in the crossover region, with the maximum of $T_c$ occurring at $E_{c1}$.\cite{Park2019prb,supplement} Phonon spectra in the superconducting state $\rho_{ph}^{sc}(\omega)$ depict the development of collective modes as sharp peaks coupled with soft phonon modes starting from the vicinity of the crossover regime, $E_b=0.8$ (black curve), which is smaller than $E_{c1}$ as shown in Fig. 1 (b). Compared to the soft mode, which is almost tied to the zero frequency at $E_{c2}$ in the normal state, the collective mode at $E_b=2.2$ in the superconducting state is slightly shifted above zero frequency. Even for the strong coupling (BEC) regime, $E_b=2.6 > E_{c2}$ (green curve), corresponding to the insulating region of the normal state, the weight of the soft collective mode remains nonvanishing due to phase fluctuation stemming from the formation of local pairs.

The density-density correlation spectral function in the superconducting state is derived from $\rho_{c}^{sc}(\omega)= -1/\pi \textrm{Im} \chi_c(\omega)$ where $\chi_c (\omega)=\llangle n_f, n_f \rrangle_\omega$, and is compared to the phonon spectrum in Fig. 2. 
As depicted in Fig. 2 (a), since the density-density correlation spectra provide information about collective modes in the low-frequency region, the phonon spectra in Fig. 2 (b) also cover the same frequency range by zooming in Fig. 1 (b). This allows us to confirm the consistent evolution of collective modes between the density-density correlation and the phonon spectra. 
In the weak coupling BCS regime ($E_b=0.4$), the bosonic spectra exhibit a small and broad peak or shoulder around $\omega= 2 \Delta_P$ as a gap mode.
This gap mode arises due to the many-body states indirectly influenced by coherent pairs in the absence of long-range interaction and may be too tiny to observe experimentally.
It is important to note that the gap mode should be distinguished from the amplitude mode, which does not directly couple to the charge in linear response, as pointed out by Littlewood and Varma.\cite{Littlewood_PRL1981,Littlewood_PRB1982}

For $E_b \ge 0.8$, the peak position of collective mode ($\omega_{col}$) follows the soft phonon mode ($\omega_s$) less than $2\Delta_P$, while the peak height significantly increases and the gap mode becomes less prominent. 
We present the evolution of the collective modes in Fig. 3 (a). The pairing gap $\Delta_P$ and the superfluid stiffness $D_s$  are taken from Ref. \onlinecite{Park2019prb}. Detailed calculations of them are explained in the Supplemental Material (SM III).\cite{supplement}. In the BCS regime ($E_b < 0.6$), $\omega_{col}$ is proportional to $2 \Delta_P$, while from the crossover regime ($0.6 < E_b < 1.3$, less than $E_{c1}$ where $\omega_s < 2 \Delta_P$), $\omega_{col}$ follows $\omega_{s}$. In the BEC regime for $E_b > E_{c2}$, where $\omega_s$ disappears, $\omega_{col}$ scales with the weak superfluid stiffness $D_s$.  Furthermore, as the coupling goes to infinity $(E_b \rightarrow \infty)$, this collective mode approaches zero, corresponding to the massless Goldstone mode. We also estimated the weight of the collective mode as
$W = \int_0^{2 \omega_{col}} \rho_{ph}^{sc} (\omega)$  in Fig. 3 (b). It turns out that $W$ is proportional to the magnitude of the pairing $\Delta_P$ for the entire coupling range,
indicating that the pairing order parameter $\Delta_P$ emerges due to the U(1) gauge symmetry breaking.

Therefore, the evolution of the collective modes from the gap mode to the soft boson mode due to increased phase fluctuation can be understood based on variations in pairing coherence and the ground state of the normal state.
The BCS regime corresponds to the metallic ground state of the normal state, where the introduction of superconducting pairing leads to a pairing coherence with strong superfluid stiffness, resulting in the emergence of the pairing gap mode in the bosonic spectra.
On the other hand, the BEC regime corresponds to the bipolaron insulating state of the normal state, characterized by ground state degeneracy between empty and doubly occupied states. 
In this composite boson system, the introduction of superconducting pairing transforms the ground state into singlet bound pairs with a significant loss of the pairing coherence, leading to large phase fluctuations and the substantial development of the phase mode.

\begin{figure}
\begin{center}
\includegraphics[width=\linewidth]{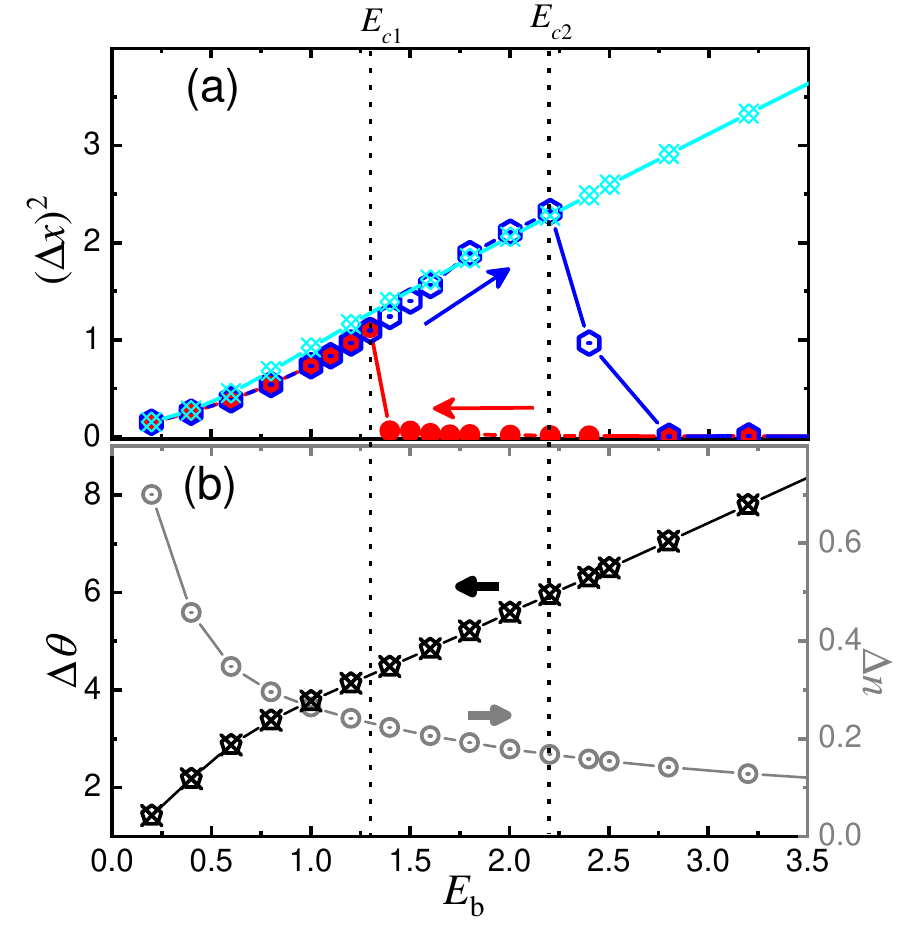}
 \caption{(a) Square of the lattice displacement fluctuation $\left( \Delta x \right)^2$ as a function of  $E_b$ for the normal state with metallic initial configuration (blue data) and with insulating one (red data) and for the superconducting state (cyan data). Blue and red arrows indicate the direction of the calculation procedure from small $E_b$ (using the metallic initial configuration) and large $E_b$ (using the insulating one), respectively. (b) Phase fluctuation (black data) and the fluctuation of the particle number (grey data) as a function of $E_b$. $E_{c1}$ and $E_{c2}$ are noted as dotted lines.}
\label{fig:exp_fluc}
\end{center}
\end{figure}

{\it Expectation values and the phase fluctuation }\label{sec:resultsB}
--
We have examined the evolution of the collective phase mode across the BCS-BEC crossover and now estimate the phase fluctuation $\Delta \theta$ based on the uncertainty relation.  Based on the mean-field solution of the Hamiltonian in Eq.\ (\ref{eq:Hol}), we establish the relationship between the number of particles and the lattice displacement as $\left< n-1 \right> = (\omega_0 / g ) \left<x \right> / 2 $ because they are linearly coupled in Eq.\ (\ref{eq:Hol}). The expectation values are computed from the ground state in both the normal and superconducting state, confirming the equality as shown in the Supplemental Material (SM IV).\cite{supplement} The quantum fluctuation of the lattice displacement $\Delta x$ at zero temperature can be calculated from $(\Delta x)^2  \sim \left< x^2 \right> - \left< x \right>^2$ and we plot the square of lattice fluctuation for the normal and superconducting states in Fig. 4 (a). In the normal state, the lattice fluctuation increases with increasing the coupling for the metallic state but drops to zero for the insulating state, illustrating the hysteresis loop between $E_{c1}$ and $E_{c2}$. However, in the superconducting state, the lattice fluctuation gradually increases with the coupling through $E_{c1}$ and $E_{c2}$. From the relation between the number of particles and the lattice displacement, we can calculate the fluctuation of the number of particles as $\Delta n = \omega_0 \Delta x / (2g)$ and the phase fluctuation $\Delta \theta \sim 1/\Delta n $ from the uncertainty relation. We present the results of  $\Delta \theta$ and  $\Delta n$ in Fig. 4 (b) and confirm that the phase fluctuation increases with increasing coupling.

$Conclusion$\label{sec:con}
-- 
We investigated the collective mode with the phase fluctuation across the BCS-BEC crossover, emerging in bosonic spectra for the Holstein model within the superconducting state, employing DMFT-NRG calculation. Phonon spectra and the density-density correlation functions were computed to examine the collective mode associated with U(1) gauge symmetry breaking. Consistent results were observed between these spectra, revealing the evolution of collective excitations. In the weak coupling (BCS) regime, the collective mode appears near $\omega_{col} \sim 2 \Delta_P$ and increases with coupling strength. Transitioning to the intermediate coupling (crossover) regime, where the soft phonon mode, caused by the lattice instability, dropped below $2 \Delta_P$, the collective mode tracked the soft phonon mode ($\omega_{col} \sim \omega_s$) and decreased with increasing coupling. In the strong coupling (BEC) regime, the collective mode decreased proportionally to the superfluid stiffness ($\omega_{col} \sim D_s/5$) with increasing coupling. This collective mode stemmed from increasing phase fluctuation in local pairs. The weight of the collective mode was found to be proportional to the pairing gap amplitude, reflecting U(1) gauge symmetry breaking across the entire coupling regime. 

We anticipate that this collective mode can be experimentally observed through techniques such as inelastic X-ray scattering, Raman spectroscopy, and electron energy loss spectroscopy in systems like magic-angle twisted bilayer and trilayer graphene\cite{Oh_Nature2021,Park_Nature2021}, as well as iron chalcogenides\cite{Mizukami_CP2023} superconductors, which exhibit the BCS-BEC crossover.\cite{Chen_RMP2024} While our analysis focused on antiadiabatic phonons, we have confirmed that the findings are also valid for adiabatic phonons (not shown in this paper). For instance, the bosonic spectral evolution for $\omega_0=0.05$ ($ \ll t$) reveals a non-monotonic variation in the frequency of the collective mode, driven by increasing phase fluctuations throughout the BCS-BEC crossover. As a universal feature of U(1) gauge symmetry breaking, the weight of the collective mode remains proportional to the gap amplitude, even in the adiabatic phonon regime. Future work will focus on examining pair-pair correlation functions and optical conductivity to further explore the amplitude mode.


\begin{acknowledgments}

We acknowledge the support from National Research Foundation of Korea under NRF-RS-2023-00246909 (THP) and NRF-2021R1F1A1063697 (HYC).

\end{acknowledgments}


\bibliography{references.bib}{}

\end{document}


\setcounter{equation}{0}
\setcounter{figure}{0}
\setcounter{table}{0}
\setcounter{page}{1}
\makeatletter
\renewcommand{\theequation}{S\arabic{equation}}
\renewcommand{\thefigure}{S\arabic{figure}}
\renewcommand{\bibnumfmt}[1]{[S#1]}
\renewcommand{\citenumfont}[1]{S#1}

\title{Supplemental Material for \\ Collective mode across the BCS-BEC crossover in Holstein model }

\author{Tae-Ho Park}
\affiliation{Department of Physics and Institute for Basic Science Research, Sungkyunkwan University, Suwon 16419,
Korea}

\author{Han-Yong Choi}
\affiliation{Department of Physics and Institute for Basic Science Research, Sungkyunkwan University, Suwon 16419,
Korea}
\affiliation{Asia Pacific Center for Theoretical Physics, Pohang 37673, Korea}





\maketitle

\onecolumngrid

\vspace{2cm}

{\bf

S1. Derivation of effective electron-electron interaction from electron-phonon coupling \\ \\

S2. Phonon spectral function and density-density correlation function \\ \\

S3. Superfluid stiffness, pairing gap and superconducting critical temperature \\ \\

S4. Expectation values \\ \\

}

\newpage

\section{S1. Derivation of effective electron-electron interaction from electron-phonon coupling}\label{}

The Hamiltonian for the Holstein model, Eq.(1) in the main text, can be reformulated in terms of the kinetic and potential energies of phonon as follows:
\begin{equation}\label{eq:Holph}
{\cal H} = -\sum_{\langle i,j\rangle\sigma}  t_{ij}c_{i\sigma}^\dag c_{j\sigma}  - \mu \sum_i n_i 
+\frac{1}{2 \omega_0} \sum_i \left[ p_i^2 + \omega_0^2 x_i^2 \right]
+ g \sum_i  x_i \left( n_i -1 \right),
\end{equation}
where $p_i$ represents the phonon momentum at site $i$. Using the Lagrangian formalism, $p_i$ can be replaced by the time derivative of the displacement, $p_i = \partial_t x_i$, assuming the mass $m=1$.
The partition function $Z$ can be written as a path integral over the Grassmann variables $c^*$ and $c$ for the electronic operators, and  over the phonon field $x$:
\begin{equation}\label{eq:partition}
Z=\textrm{Tr} e^{-\beta {\cal H}}=\int D [x] D [c^*,  c] e^{-S\left[ c^*, c, x \right]}
\end{equation}
where $\beta = 1/T$ and $T$ is the temperature. The action of the system in imaginary time $\tau$ is then given by
\begin{equation}\label{eq:action}
S\left[ c^*, c, x \right] = \int_0^{\beta} d \tau \left[  \sum_{i \sigma} c_{i\sigma}^* (\partial_{\tau} -\mu) c_{i \sigma}  -\sum_{\langle i,j\rangle\sigma}  t_{ij}c_{i\sigma}^* c_{j\sigma}
+ \frac{1}{2 \omega_0}  \sum_{i}  \left[|\partial_{\tau} x_i|^2 + \omega_0^2 x_i^2 \right]  + g \sum_i  x_i \left( n_i -1 \right) \right] .
\end{equation}

Within dynamical mean-field theory (DMFT), we solve an impurity problem for which the effective action at the impurity site is
\begin{equation}\label{eq:effaction}
S_{eff}= -\int_0^{\beta} d \tau \int_0^{\beta} d \tau'  \sum_{\sigma} c_{\sigma}^* (\tau) {\cal G}_0^{-1}(\tau - \tau' ) c_{\sigma} (\tau') + \frac{1}{2 \omega_0} \int_{0}^{\beta} d \tau \left[ |\partial_{\tau} x|^2 + \omega_0^2 x^2 (\tau) \right]
+ g \int_0^{\beta} d \tau x(\tau) (n(\tau)-1)
 \end{equation}
 where ${\cal G}_0(\tau - \tau' )$ is the Weiss effective field.\cite{georges_1996}
 
 To express the phonon field in Matsubara frequency $\omega_n \equiv 2n\pi T$ (with integer $n$), we use the Fourier transformation:
 \begin{equation}\label{eq:fourier}
 x(i \omega_n) = \int_0^{\beta} d \tau e^{i \omega_n \tau} x(\tau) , \  \  \  x(\tau)= \frac{1}{\beta} \sum_{\omega_n} e^{-i \omega_n \tau} x(i \omega_n).
 \end{equation}
The free phonon action $S_{ph}$, which is the second term of Eq.\ (\ref{eq:effaction}), can be rewritten in terms of the bare phonon propagator as
 \begin{equation}\label{eq:phaction}
S_{ph}= -\frac{1}{\beta^2} \sum_{\omega_n} D_0^{-1}(i \omega_n) x^2(i \omega_n) = -\int_0^{\beta} d \tau \int_0^{\beta} d \tau' x(\tau) D_0^{-1}(\tau) x(\tau')
 \end{equation} 
 where the bare phonon Green's functions in Matubara frequency and imaginary time are respectively
 \ba
 &D_0&(i \omega_n) = - \frac{2 \omega_0}{\omega_n^2 + \omega_0^2},
 \\
 &D_0&(\tau) = \frac{e^{-\omega_0 |\tau|}}{1-e^{-\omega_0 \beta}} + \frac{e^{\omega_0 |\tau|}}{e^{\omega_0 \beta}-1}.
 \ea 
Then, integrating out the phonon variable $x$ via Gaussian integration in the effective action Eq.\ (\ref{eq:effaction}), we obtain
\begin{equation}\label{eq:Hubbard}
S_{eff}= -\int_0^{\beta} d \tau \int_0^{\beta} d \tau'  \left[ \sum_{\sigma} c_{\sigma}^* (\tau) {\cal G}_0^{-1}(\tau - \tau' ) c_{\sigma} (\tau') 
+  (n(\tau)-1) U_{eff}(\tau - \tau') (n(\tau')-1) \right].
 \end{equation}
 This effective action is similar to that of the Hubbard model, where the on-site electron-electron interaction is replaced by a retarded interaction due to electron-phonon coupling. 
 Here, $ U_{eff}(\tau) = g^2 D_0(\tau)$, and in real frequency, the effective interaction becomes 
 \begin{equation}\label{eq:fourier}
 U_{eff}(\omega) = g^2 \frac{2 \omega_0} {(\omega^2 - \omega_0^2)}.
  \end{equation}
  
\section{S2. Phonon spectral function and density-density correlation function}\label{}

Here, we describe the calculation of bosonic spectra for the phonon Green’s function and the density-density correlation function.
The phonon propagator and the density-density correlation function on the local site are defined, respectively, by
\ba
D(\omega)=\llangle x,x \rrangle_\omega ,  ~~~  \chi_c (\omega)=\llangle n_f, n_f \rrangle_\omega,
\ea
where $x (=a+a^\dag)$ is the local lattice deformation operator and $n_f (=\sum_\sigma f_\sigma^\dag f_\sigma)$ is the electron density operator at a local site from the Hamiltonian of Eq.(2a) in the main text. The double bracket $\llangle \rrangle$ represents the correlator defined as:
 \ba
\llangle {\cal O}_1,{\cal O}_2 \rrangle_\omega &=&
\int_{-\infty}^{\infty} dt\ e^{i\omega t}\ \llangle {\cal
O}_1,{\cal O}_2 \rrangle_t ,
 \\ \nonumber
\llangle {\cal O}_1,{\cal O}_2 \rrangle_t &=& -i \theta(t) \left<
[{\cal O}_1 (t),{\cal O}_2 (0)] \right>=-i \theta(t) \left<
[{\cal O}_1 (0),{\cal O}_2 (-t)] \right>,
 \ea
where ${\cal O}_1$ and ${\cal O}_2$ are arbitrary boson operators, $\theta$ is the step function, $\left<~~\right>$ is the
thermodynamic average, and $\left[~~\right]$ denotes the boson commutator.\cite{jeon_2003}
The equation of motion for the correlator is given by
 \ba
 \label{eom}
\omega \llangle {\cal O}_1,{\cal O}_2 \rrangle_\omega = \left< [
{\cal O}_1,{\cal O}_2 ] \right> + \llangle [{\cal
O}_1,{\cal H} ] , {\cal O}_2 \rrangle_\omega
= \left< [ {\cal O}_1,{\cal O}_2 ] \right> - \llangle
{\cal O}_1 , [{\cal O}_2,{\cal H} ] \rrangle_\omega ,
 \ea
 where $\cal H$ is the Hamiltonian of the system.
 By inserting the boson operator $x$ into ${\cal O}_1$ and ${\cal O}_2$ and replacing $\cal H$ with Eq.(2a) of the main text in Eq.\ (\ref{eom}), we obtain 
 \ba
 \left( \omega^2 - \omega_0^2 \right) \llangle x,x \rrangle_\omega = 2 \omega_0  \left( 1 + g  \llangle n_f,x \rrangle_\omega \right)
 \ea
which leads to the Dyson equation form for the phonon Green's function as
\ba
\label{dyson}
D^{-1}(\omega) &=& D_0^{-1}(\omega)-\Pi(\omega),
\\ \nonumber
D_0(\omega) = \frac{2\omega_0}{(\omega^2-\omega_0^2)}
&,&
\Pi(\omega) = g\frac{\llangle n_f, x\rrangle_{\omega}}{\llangle  x, x \rrangle_{\omega}},
\ea
where $D_0(\omega)$ is the bare phonon Green's function and $\Pi(\omega)$ is the phonon self-energy.
The sum rule that $D(\omega)$ satisfies is:
\begin{equation}\label{sumrule}
-\frac{1}{\pi} \int_{-\infty}^{\infty} d\omega \textrm{Im} D(\omega) n_B(\omega) = \left< x^2 \right> ,
\end{equation}
where $n_B(\omega)$ is the Bose function.
In the spectral form, the imaginary part of correlators in Eq.\ (\ref{dyson}) are calculated using the Lehmann representation as
\ba \label{eq:ImPG}
-\frac{1}{\pi} \textrm{Im} \llangle  x, x \rrangle_{\omega} &=& \frac {1}{Z} \sum_{n m} \lvert  \langle  n \lvert  x \rvert m \rangle \rvert^2 \delta ( \omega - (E_n -E_m))(e^{-\beta E_n} - e^{-\beta E_m}),
\\ \label{eq:ImSE}
-\frac{1}{\pi} \textrm{Im} \llangle  n_f, x \rrangle_{\omega} &=& \frac {1}{Z} \sum_{n m} \lvert  \langle  n \lvert  n_f\rvert m \rangle  \langle  m \lvert  x \rvert n \rangle \rvert \delta ( \omega - (E_n -E_m))(e^{-\beta E_n} - e^{-\beta E_m}),
\ea
and the spectral function of density-density correlation function is
\begin{equation} \label{eq:ImDen}
\rho_c(\omega)=-\frac{1}{\pi} \textrm{Im} \chi_c(\omega) = \frac {1}{Z} \sum_{n m} \lvert  \langle  n \lvert  n_f \rvert m \rangle \rvert^2 \delta ( \omega - (E_n -E_m))(e^{-\beta E_n} - e^{-\beta E_m}),
\end{equation}
where $Z$ is the partition function, $\beta= 1/k_B T$ (with $k_B$ being the Boltzmann constant and $T$ the temperature). $E_n$ and $E_m$ are eigenenergies obtained from the numerical renormalization group (NRG) calculation, and  $\lvert n \rangle = \lvert Q, S, w_n \rangle $ and $\lvert m \rangle = \lvert Q, S, w_m \rangle $ represent the corresponding eigenstates, labeled by quantum numbers: the charge $Q$, the total spin $S$ and $w_n$ ($w_m$) the labeling number for the diagonalized states on a $Q$ and $S$ block matrix for the normal state.\cite{wilson_1975,krishnamurthy_1980} In the superconducting state, the eigenstates are represented by the spin quantum number as $\lvert n \rangle = \lvert S, w_n \rangle $ and $\lvert m \rangle = \lvert S, w_m \rangle $ without charge $Q$.
The real part of correlators is obtained from the Kramers-Kronig relation as
\ba
\label{eq:KK}
\textrm{Re} \llangle  x, x \rrangle_{\omega} &=& P \int_{-\infty}^{\infty} \frac{d \omega '}{\pi} \frac{\textrm{Im} \llangle  x, x \rrangle_{\omega'}}{\omega ' - \omega},
\\
\textrm{Re} \llangle  n_f, x \rrangle_{\omega} &=& P \int_{-\infty}^{\infty} \frac{d \omega '}{\pi} \frac{\textrm{Im} \llangle  n_f, x \rrangle_{\omega'}}{\omega ' - \omega}.
\ea
Finally, the phonon spectral function is calculated from the Eq.\ (\ref{dyson}) as
\begin{equation}
\rho_{ph}(\omega)=-\frac{1}{\pi} \textrm{Im} D(\omega).
\end{equation}

\section{S3. Superfluid stifffness ($D_s$), pairing gap ($\Delta_P$) and superconducting critical temperature ($T_c$) }\label{}

The superfluid stiffness $D_s$ is a critical parameter related to the phase coherence of the superconducting phase and indicates the energy cost of phase fluctuations.\cite{toschi_2005,bauer_2009,grag_2005}
$D_s$ can be measured or calculated from the zero-frequency weight of the real part of the optical conductivity, and its the upper bound 
is represented by the ratio of the superfluid density ($n_s$) to the effective mass ($m^*$), derivable from the kinetic energy within the BCS theory. 
Experimentally, this is accessible via the inverse square of the magnetic field penetration depth $\lambda$ ($\lambda^{-2}= (4 \pi n_s e^2)/(c^2m^*)$ in CGS units, where $e$ is the charge of an electron and $c$ is the speed of light),
derived from the Meissner effect.\cite{Benfatto_PRB2001,Ghosal_PRB2001}
We calculate $D_s$ based on the Kubo formula as\cite{Scalapino_PRB1993,Scalapino_PRL1992}
\begin{equation}\label{eq:DsKubo}
D_s = -\left< E_{kin}\right> -\Lambda_T \left( \mathbf{q} \rightarrow 0, i\omega_n = 0 \right).
\end{equation}
In the first term, $\left< E_{kin}\right>$ is the kinetic energy calculated from the Hamiltonian Eq.(1) in the main text as 
\ba
\label{eq:Kin}
\left< E_{kin}\right> &=& - \frac{2 t}{N_s} \sum_{\langle i,j\rangle\sigma} \left< c_{i\sigma}^\dag c_{j\sigma} \right> = \sum_{k} \varepsilon_k \left<n_k \right>
\\ \nonumber
&=& \frac{2}{\beta} \sum_{n} \int d\varepsilon_k   D(\varepsilon_k)\varepsilon_k G_k(i\omega_n),
\ea
where $N_s$ is the number of sites, and $D(\varepsilon_k)=\sqrt{4t^2-\varepsilon_k^2}/(2 t^2 \pi)$ is the density of states of noninteracting electrons in a semicircular form for Bethe lattice, considering the infinite dimension of DMFT calculation. $G_k(i\omega_n)$ is the diagonal (1,1) component of the lattice Green's function $\hat{\mathbf{G}}_k(i\omega_n)$ in the $2\times2$ Nambu matrix notation in the presence of superconductivity in Matsubara frequency $ i\omega_n$ space.\cite{Park2019prb}
The term $-\left< E_{kin}\right>$ represents the diamagnetic response to an external magnetic field. 
Meanwhile, the paramagnetic response is given by the transverse current-current correlation in momentum $\mathbf{q}$ and  $ i\omega_n$ space as
\begin{equation}\label{eq:CurCor}
\Lambda_T \left( \mathbf{q} , i\omega_n \right) = \frac{1}{N} \int_0^{\beta} d \tau e^{i \omega_n \tau} \left< j_T^p (\mathbf{q},\tau) j_T^p (-\mathbf{q},0) \right>
\end{equation}
with the transverse component of the paramagnetic current $ j_T^p (\mathbf{q},\tau) $. Then, the second term in Eq.\ (\ref{eq:DsKubo}) becomes
\begin{equation}\label{eq:Para}
\Lambda_T \left( \mathbf{q} \rightarrow 0, i\omega_n = 0 \right) = -\frac{2}{\beta} \sum_{n}\int d \varepsilon_k D(\varepsilon_k)V(\varepsilon_k) 
\left[ G_k (i\omega_n)G_k^* (i\omega_n) + G_k^{\textrm{off}}(i\omega_n) G_k^{\textrm{off}}(i\omega_n)\right],
\end{equation}
where $V(\varepsilon_k)=(4t^2-\varepsilon_k^2)/3$ and $G_k^{\textrm{off}}(i\omega_n)$ is the off-diagonal component of $\hat{\mathbf{G}}_k(i\omega_n)$. Using the relation $\partial /\partial \varepsilon_k \left[ D(\varepsilon_k)V(\varepsilon_k) \right] = - \varepsilon_k D(\varepsilon_k) $ and integration by part to evaluate Eq.\ (\ref{eq:DsKubo}), 
we obtain 
\begin{equation}\label{eq:SFstiffness}
D_s=\frac{4}{\beta} \sum_n \int d\varepsilon_k D(\varepsilon_k)V(\varepsilon_k)  G_k^{\textrm{off}}(i\omega_n) G_k^{\textrm{off}}(i\omega_n),
\end{equation}
where the diamagnetic term is canceled in the DMFT calculation based on the Bethe lattice. In terms of the analytic continuation, $i\omega_n \rightarrow \omega + i \delta$, 
the lattice Green's function in real frequency space is represented as
\ba\label{eq:GFreal}
\hat{\mathbf{G}}_k^{-1} (\omega) &=&\omega \tau_0-(\varepsilon_k-\mu)  \tau_3-\hat{\mathbf{\Sigma}}(\omega) 
\\ \nonumber
&=& W(\omega) \tau_0 - (\varepsilon_k - \mu) \tau_3 - \phi(\omega)\tau_1,
\ea
where  $\tau_i$ ($i=1,2,3$) are the Pauli matrices, $\tau_0$ is the $2 \times 2$ identity matrix, and $\mu$ is the chemical potential.
Here, we decomposed the self-energy matrix  to
$\hat{\mathbf{\Sigma}}( \omega )=\Sigma (\omega) \tau_0+\phi (\omega) \tau_1$
and  defined a function $W(\omega) \equiv \omega-\Sigma (\omega)$ for the diagonal component.
Inserting the spectral representation:
\begin{equation}\label{eq:SpecRep}
G_k^{\textrm{off}}(i\omega_n) = -\frac{1}{\pi} \int_{-\infty}^{\infty} d \epsilon \frac{\textrm{Im}G_k^{\textrm{off}}(\epsilon)}{i \omega_n -\epsilon}
\end{equation}
into Eq. \ (\ref{eq:SFstiffness}), we can represent $D_s$ as
\begin{equation}\label{eq:SFstiffness_w}
D_s=-\frac{8}{\pi} \int d\varepsilon_k D(\varepsilon_k)V(\varepsilon_k) \int_{-\infty}^0 d\omega \textrm{Im}G_k^{\textrm{off}}(\omega) \textrm{Re}G_k^{\textrm{off}}(\omega),
\end{equation}
with
\begin{equation}\label{eq:Goff}
 G_k^{\textrm{off}}(\omega) 
=\frac{\phi(\omega)}{W(\omega)^2-(\varepsilon_k -\mu) ^2-\phi(\omega)^2}.
\end{equation}
Inserting Eq.\ (\ref{eq:Goff}) into Eq.\ (\ref{eq:SFstiffness_w}), we obtain
\begin{equation}\label{eq:stiffness}
D_s=-\frac{2}{3\pi t^2}  \int_{-\infty}^0 d\omega \left[ 1- \frac{\sqrt{a(\omega)^2-1}}{a(\omega)} -\frac{\sqrt{a(\omega)^2-1}}{2a(\omega)^3} \right],
\end{equation}
where $a(\omega)=\sqrt{W(\omega)^2-\phi(\omega)^2}/(2t) $.
For the Holstein model (Eq.\ (1) of the main text), the chemical potential is equal to zero at half-filling due to the particle-hole symmetriy about $\mu=0$.
The upper bound of the superfluid stiffness within BCS theory is $D_s= \pi n_s e^2 / m^*$.
In the absence of the superconductivity, the $\mathbf{q}=0, \omega \rightarrow 0 $ limit of the optical conductivity determines the charge stiffness or Drude weight $D_n$,
similar form with Eq.\ (\ref{eq:DsKubo}) from the Kubo formula
\begin{equation}\label{eq:DnKubo}
D_n = -\left< E_{kin}\right> -\textrm{Re}\Lambda_T \left( \mathbf{q} = 0, \omega \rightarrow 0 \right),
\end{equation}
and $D_n= \pi n e^2 / m^*$ where $n$ is the density of mobile charge carriers in a normal metal.

We show the calculated $D_s/2$ as a function of $E_b$ in Fig. S1 with the pairing gap $\Delta_p /1.75$,
scaled with the superconducting critical temperature $T_c$ in BEC strong coupling regime and BCS weak coupling regime, respectively.
We obtain the gap function by renormalization of the off-diagonal self-energy as
$\Delta(\omega)=\phi(\omega)/Z(\omega)$,
where the renormalization function is $Z(\omega)=1-\Sigma(\omega)/\omega$.
The pairing gap $ \Delta_p$ is given by the solution of 
$ \Delta_p= \textrm{Re}\Delta(\omega = \Delta_p)$. This procedure is illustrated in Ref. \onlinecite{Park2019prb}.
The solutions graphically represent the intersection points of $\textrm{Re} \Delta (\omega)$
and the linear line ($y=\omega $).
The superconducting critical temperature $T_c$ is obtained from the exponentially diverging energy flow approaching the SC phase fixed point, indicating an instability of the normal phase in the NRG-DMFT iteration procedure \cite{hecht_2008,kuleeva_2014}.
The renormalization group flow is presented by the variation of the NRG site number $N$. 
The $T_c$ is determined by $ T_c = \Lambda^{-N_{T_c}/2}$, where $N_{T_c} \sim -2 \log y_0/\log \Lambda$
and $y_0$ is the y-axis intercept of the energy flow.

\begin{figure}[h]
\begin{center}
\includegraphics[width=0.6\linewidth]{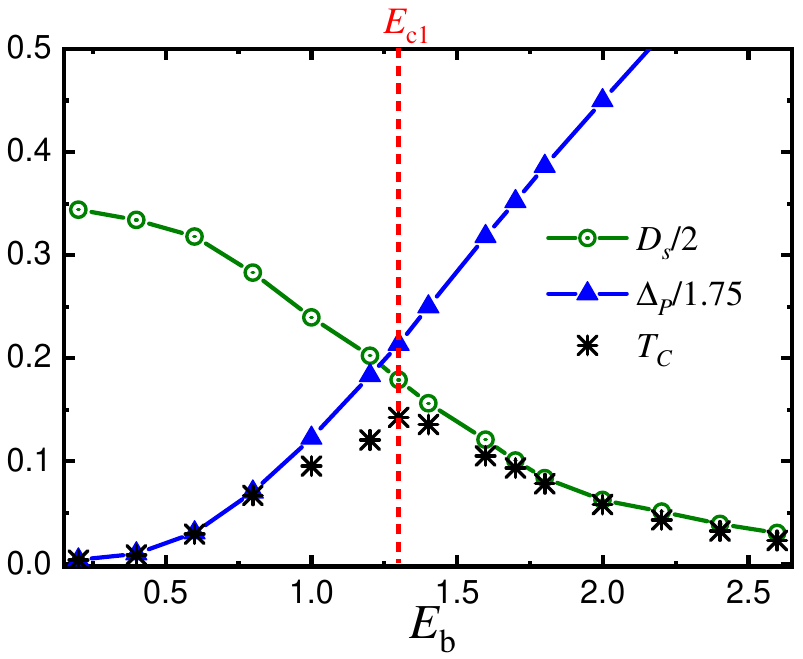}
 \caption{NRG-DMFT calculation results of the superfluid stiffness $D_s/2$ (green circle), the pairing gap $\Delta_p/1.75$ (blue triangle), and the superconducting critical temperature $T_c$ (black asterisk) as a function of $E_b$. The red vertical dashed line indicates the critical value $E_{c1}=1.3$ for the metal-insulator transition in the normal state and the maximum $T_c$ lies on $E_{c1}$. We use the unit of energy with the half-bandwidth $D$. }
\label{fig:energy}
\end{center}
\end{figure}

\section{S4. Expectation values }\label{}

Based on the mean-field solution of the Hamiltonian given in Eq.(1) in the main text, the relationship between the number of particles and the lattice displacement can be expressed as  $\left< n-1 \right> = (\omega_0 / g ) \left<x \right> / 2 $. The expectation values are calculated from the ground state in both the normal and the superconducting state, confirming the equality as shown in Fig. S2. In the normal state, we considered two different initial conditions, revealing phase separation between $E_{c1}$ and $E_{c2}$. For the metallic state, $\left< n \right>=1$. However, in the bipolaron insulating state, $\left< n \right> -1 \approx 1$ or $-1$ for the doubly occupied and empty states, respectively. In contrast, for the superconducting state, we observe a metal-like unique solution with $\vert \left< n - 1 \right> \vert \sim 0$ across the entire coupling range, regardless of the initial configurations.

\begin{figure}[h]
\begin{center}
\includegraphics[width=0.7\linewidth]{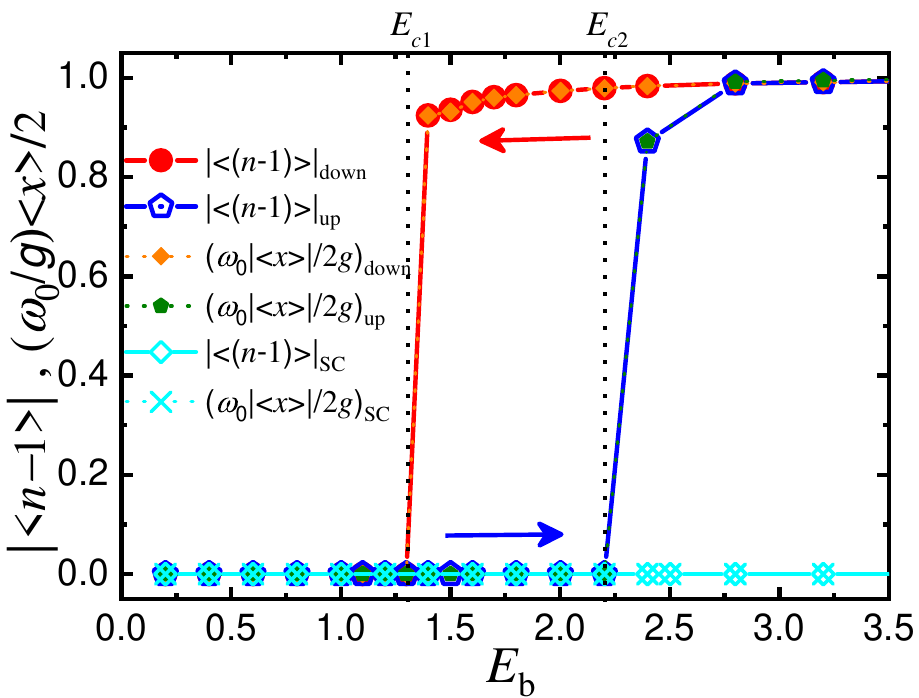}
 \caption{ Expectation values of $\vert \left< n-1 \right> \vert$ and $ (\omega_0 / g ) \left<x \right> / 2 $ as a function of  $E_b$ for the normal state with metallic initial configuration (blue and green data) and with insulating one (red and orange data) and for the superconducting state (cyan rhombus and cross data). The black vertical dotted lines indicate the critical values $E_{c1}=1.3$ and $E_{c2}=2.2$ for the metal-insulator transition in the normal state. Blue and red arrows indicate the direction of the calculation procedure from small $E_b$ to up (using the metallic initial configuration) and from large $E_b$ to down (using the insulating one), respectively.}
\label{fig:expectation}
\end{center}
\end{figure}

\vspace{0.2cm}

\bibliography{references.bib}{}